\documentclass[prl,twocolumn,showpacs,preprintnumbers,amsmath,amssymb, superscriptaddress]{revtex4-2}

\usepackage{amsmath}   
\usepackage{amssymb}
\usepackage{graphicx}
\usepackage{hyperref}
\usepackage{fixmath}

\hypersetup{colorlinks,linkcolor=blue,urlcolor=blue,citecolor=blue}

\newcommand{\avg}[1]{\langle #1\rangle}

\def\e{\varepsilon}
\def\tr{\textrm{tr}}
\def\pd{\partial}

\begin{document}
\author{Doru Sticlet}
\email{doru.sticlet@itim-cj.ro}
\affiliation{National Institute for R\&D of Isotopic and Molecular Technologies, 67-103 Donat, 400293 Cluj-Napoca, Romania}
\author{Bal\'azs D\'ora}
\affiliation{Department of Theoretical Physics and MTA-BME Lend\"ulet Topology and Correlation Research Group, Budapest University of Technology and Economics, 1521 Budapest, Hungary}
\author{C\u{a}t\u{a}lin Pa\c{s}cu Moca}
\affiliation{MTA-BME Quantum Dynamics and Correlations Research Group, Institute of Physics, Budapest University of Technology and Economics, 1521 Budapest, Hungary}
\affiliation{Department  of  Physics,  University  of  Oradea,  410087,  Oradea,  Romania}
\noaffiliation

\title{Kubo Formula for Non-Hermitian Systems and Tachyon Optical Conductivity}

\begin{abstract}
	Linear response theory plays a prominent role in various fields of physics and provides us with extensive information about the thermodynamics and dynamics of quantum and classical systems.
	Here we develop a general theory for the  linear response in non-Hermitian systems with nonunitary dynamics and derive a modified Kubo formula for the generalized susceptibility for an
	arbitrary (Hermitian and non-Hermitian) system and perturbation.
	We use this to evaluate the dynamical response of a non-Hermitian, one-dimensional Dirac model with imaginary and real masses, perturbed by a time-dependent electric field. 
	The model has a rich phase diagram, and in particular, features a tachyon phase, where excitations travel faster than an effective speed of light.
	Surprisingly, we find that the dc conductivity of tachyons is finite, and the optical sum rule is exactly satisfied for all masses.
	Our results highlight the peculiar properties of the Kubo formula for non-Hermitian systems and are applicable for a large variety of settings. 
\end{abstract}
\maketitle

\paragraph{Introduction.}Linear response theory is a fundamental framework in many branches of physics that describes the way in which a physical system responds to small external perturbations and allows us to interpret physical measurements~\cite{Mahan2000}.
The response is expressed in terms of a generalized susceptibility of the unperturbed system. 
The change in the expectation value of an operator $A$ due to a time-dependent perturbation $V(t) = B f(t)$ is assumed to be linear in the perturbative force $f(t)$, 
\begin{equation}\label{eq:definition_chi}
	\delta\avg{A(t)} = \int dt'\chi_{AB}(t,t')f(t')\, ,
\end{equation}
with $\chi_{AB}(t,t')$ known as the response function. In general,  $\chi_{AB}(t,t')$ is real, and for systems invariant
under time translations $ \chi_{AB}(t,t')= \chi_{AB}(t-t')$. Traditionally, the explicit expression for the response function 
in Eq.~\eqref{eq:definition_chi} is given by the Kubo formula~\cite{Kubo1957}, and has been for decades a standard tool in the study of 
electric transport and spin susceptibilities in response to applied electric and magnetic fields in 
typical many-body~\cite{Baranger1989, Crepieux2001, Nagaosa2010}, closed~\cite{Cohen.2003,Bohr2006}, or open systems~\cite{Michel2004,Fujii2007,Kundu2009}. 
Through the fluctuation-dissipation theorem, the linear response provides information about the equilibrium fluctuations of a system, and can reveal, e.g.,~topological properties through the quantum Hall response~\cite{Thouless1982} or demonstrate the presence of fractional charge carriers for the fractional quantum Hall effect via noise measurements~\cite{Saminadayar1997}.

Recently, a newly emerging field of non-Hermitian physics~\cite{Bender2007, Rotter2009, Moiseyev2011, Cao2015, ElGanainy2018, Ashida2020,Bergholtz2021} has become very
active, mostly for its promising applications~\cite{ElGanainy2018} and for exhibiting unique features, which are absent in the Hermitian realm, such as exceptional points~\cite{Berry2004,Makris2008, Heiss2012, Zhen2015,Dora2019}, non-Hermitian skin effect~\cite{Yao2018,Kunst2018,Song2019,Li2020a,Borgnia2020}, non-Bloch phase transitions~\cite{Longhi2019}, and unidirectional invisibility~\cite{Lin2011}, to mention a few. 
By now, quantum optics and cold atoms setups provide fertile grounds where such non-Hermitian concepts are probed experimentally~\cite{Makris2008,El-Ganainy2019, Li2020b,Takasu2020}. In such state-of-the-art experiments, e.g.,~in a quantum gas microscopy experiment~\cite{Bakr2009,Kuhr2016}, the measurement backaction can qualitatively alter the low-energy physics, so it is crucial to understand and describe the effect of small perturbations in such non-Hermitian systems. 
In this regard, only recently, a non-Hermitian perturbation of a Hermitian Hamiltonian has been considered, and a modified Kubo formula for 
the generalized susceptibility has emerged~\cite{Pan2020}. Still, at present, a complete non-Hermitian linear response theory, in which the 
Hamiltonian governing the system and the perturbation are both non-Hermitian, is missing. 

In the present work we address this problem, and fill this gap,  by developing a general formalism and providing a generalized Kubo formula for the susceptibility 
$\chi_{AB}(t,t')$ introduced in Eq.~\eqref{eq:definition_chi} which is suitable for non-Hermitian systems under very general conditions. 
We use  our approach by computing the correlation function for the current operator, in response to an external electric field, and the 
associated optical conductivity for a non-Hermitian tachyon model. 
Although non-Hermiticity is in general associated with dissipation, we find remarkable features such as a finite dc conductivity in the absence of scatterers or local charge conservation, which guarantees that the  optical sum rule is satisfied~\cite{Benfatto2005}.

\paragraph{General theory.} 
We consider a general setup in which the  system  is modeled by a time-independent non-Hermitian Hamiltonian $H_0$ with a nondegenerate spectrum subjected to a time-dependent perturbation $V(t)=B f(t)$, which starts to act at $t=0$,  where $B$ is an operator which can be non-Hermitian and $f(t)$, the perturbation force, is assumed to be small and real-valued.
We are interested in the linear response of a system operator $\delta\avg{A(t)}=\avg{A(t)}-\avg{A}_0$. The expectation value $\avg{A(t)}$ at time $t$ is given by~\cite{Graefe2008}
\begin{equation}\label{avg}
	\avg{A(t)} = \frac{\text{tr}[\rho(t) A]}{\text{tr}[\rho(t)]},
\end{equation}
where $\rho(t)$ is the density matrix of the system in the Schr\" odinger representation obeying the non-Hermitian von Neumann 
equation $i\hbar\pd_t\rho(t)=H\rho(t)-\rho(t)H^\dag$, with $H=H_0+V(t)$, the full Hamiltonian. 
Its solution is decomposed as $\rho(t)=\rho_0(t)+\delta\rho(t)$ with $\rho_0(t)=e^{-iH_0t/\hbar}\rho(0)e^{iH_0^\dag t/\hbar}$, the density matrix of the unperturbed system, 
and $\delta\rho(t)$ arising from the external perturbation.
Rotating to the interaction picture (see Supplemental Material for more details), the variation $\delta\avg{A(t)}$ is evaluated to first order in the perturbation as
\begin{equation}
	\delta \avg{A(t)} = \frac{\tr[\delta\rho_I(t) A_I(t)]}{\tr[\rho_0(t)]}
	- \avg{A(t)}_0\frac{\tr[\delta\rho(t)]}{\tr[\rho_0(t)]},
\end{equation}
where $A_I(t)=e^{iH_0^\dag t/\hbar}Ae^{-iH_0t/\hbar}$ is written in the interaction representation in terms of the unperturbed system, 
and $\delta \rho_I(t) = e^{iH_0t/\hbar}\delta \rho(t)e^{-i H_0^\dag t/\hbar}$. 
The second, norm correction term, is specific to nonunitary dynamics and is absent in the unitary evolution, yet it is important to reproduce correctly the dynamics of the expectation values.
The expectation value dynamics of an operator $A$ acquires additional terms, due to the fact that $H\neq H^\dag$ and the wave function norm is not conserved. 
Keeping in mind that $A_I(t)$  is identical to the Heisenberg picture
time evolution in terms of $H_0$, i.e., $A(t)=A_I(t)$, 
we finally obtain a compact  expression for the generalized susceptibility
defined in Eq.~\eqref{eq:definition_chi} as
\begin{widetext}
	\begin{equation}\label{eq:chi_general}
		\chi_{AB}(t,t')=-\frac{i}{\hbar}\theta(\tau)
		\tr\bigg\{
		\bigg[
		[A(\tau), B]_\sim-\avg{A(t)}_0[e^{iH_0^\dag\tau/\hbar}e^{-iH_0\tau/\hbar},B]_\sim\bigg]
		\cdot \frac{\rho_0(t')}{\tr[\rho_0(t)]}\bigg\},
	\end{equation}
\end{widetext}
in terms of a modified commutator defined as
$[X,Y]_\sim = XY - Y^\dag X $, where $\tau=t-t'$. The first term is the natural 
generalization of the Hermitian Kubo formula for the non-Hermitian setting.
The second term arises solely from the nonunitary dynamics, and $\avg{A(t)}_0={\text{tr}[\rho_0(t) A]}/{\text{tr}[\rho_0(t)]}$.
In contrast to the Hermitian Kubo formula, the non-Hermitian counterpart is not time-translation invariant and depends separately on $t$ and $t'$ due to the appearance of the system density matrix $\rho_0(t)$ at various times, which may have a nonunitary evolution in the absence of the perturbation. 
Finally, note that expectation values are obtained using the right eigenvectors of the system Hamiltonian, which is a conventional approach to treat non-Hermitian systems as effective models 
of dissipative dynamics with no quantum jumps~\cite{Daley2014,Herviou2019}.
An alternative route, would be to use a biorthogonal basis~\cite{Brody2013}.
This approach is usually employed for a class of non-Hermitian Hamiltonians invariant under space-time reflection ($\mathcal{PT}$) symmetry, and endowed with a real spectrum in the $\mathcal{PT}$-symmetric phase
~\cite{Bender1998,Bender2007}.
Various $\mathcal{PT}$-symmetric Hamiltonians have been extensively studied especially in photonics by controlling gain and loss~\cite{Makris2008, Musslimani2008, Guo2009, Oezdemir2019, El-Ganainy2019}.
The eigenspectrum reality is a property shared by a larger class of pseudo-Hermitian Hamiltonians~\cite{Mostafazadeh2002, Zhang2019},
and allows us to map a non-Hermitian $\mathcal{PT}$-symmetric Hamiltonian, onto a Hermitian Hamiltonian with unitary evolution,  by using a positive definite pseudometric operator.
However, the biorthogonal approach is unsuitable for the broken $\mathcal{PT}$-symmetry phase, where 
the Hamiltonian possesses complex pairs of eigenvalues and such a mapping does not exist~\cite{Mostafazadeh2002,Mostafazadeh2007,Gardas2016}.

When the system Hamiltonian $H_0$ is Hermitian, the expression~\eqref{eq:chi_general} simplifies considerably since $\rho_0(t)=\rho(0)$ is a time-independent equilibrium density matrix. 
For example, if additionally, the perturbation proves Hermitian, the norm corrections drop out, and assuming that the system is invariant under time translation, one readily recovers the 
regular Kubo formula~\cite{Mahan2000},
$\chi(\tau)=-\frac{i}{\hbar}\theta(\tau)\avg{[A(\tau),B]}_0$. 
If the perturbation is anti-Hermitian $B^\dag=-B$, the correlation function
recovers the central result of Ref.~\cite{Pan2020}, where the modified commutator becomes an anticommutator, $\chi(\tau)=-\frac{i}{\hbar}\theta(\tau)(\avg{\{A(\tau),B\}}_0-2\avg{A}_0\avg{B}_0)$.
Finally, for a $\mathcal{PT}$-symmetric model, in the symmetric phase, or for any pseudo-Hermitian model with a real eigenspectrum, where the initial density matrix is constructed from the ground-state wave function, the correlation function Eq.~\eqref{eq:chi_general} simplifies to
\begin{eqnarray}\label{chi_ps_h}
	\chi_{AB}(\tau) &=& -\frac{i}{\hbar}\theta(\tau)\nonumber
	\avg{[A(\tau), B]_\sim -\\
		&& - \avg{A(0)}_0 [e^{iH_0^\dag \tau/\hbar}
		e^{-iH_0 \tau/\hbar},B]_\sim}_0.
\end{eqnarray}
If $H_0$ is non-Hermitian, the norm correction term is crucial and it cannot be discarded as it may encompass the dominant contribution to linear response.

\begin{figure}[t]
	\includegraphics[width=\columnwidth]{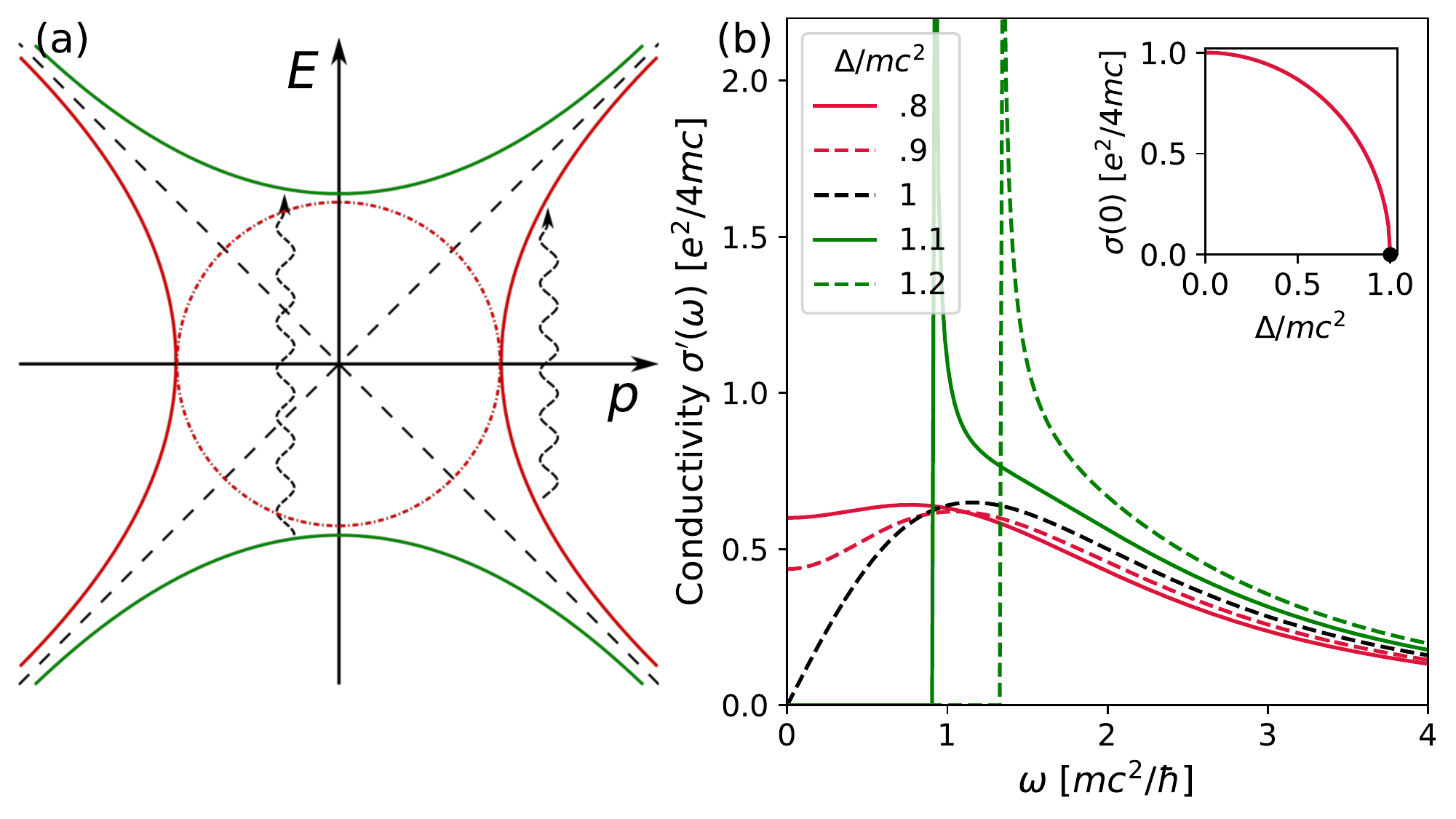}
	\caption{(a) Dispersion relation for the non-Hermitian Hamiltonian~\eqref{eq:h_tachyon} with
		(green) parabolic spectrum of gapped Dirac Hamiltonians at $|\Delta|>mc^2$,
		(red) hyperbolic tachyon dispersion for $|\Delta|<mc^2$, and (black) linear Dirac spectrum at $|\Delta|=mc^2$. The dispersion is imaginary (dashed-dotted line) only in the tachyon phase for $p^2c^2<m^2c^4-\Delta^2$.
		The wiggly lines denote resonant optical transitions under the applied electric field.
		(b) The real part of the conductivity $\sigma'(\omega)$ as a function of frequency for several values of the ratio $\Delta/mc^2$ in the case of (green) parabolic, (black) linear, or (red) hyperbolic (tachyon) dispersion.
		(Inset) Tachyon dc conductivity as a function of $\Delta/mc^2$.}
	\label{fig:1}
\end{figure}

\paragraph{Non-Hermitian Dirac Hamiltonian.}
Using our results in Eq.~\eqref{eq:chi_general}, we investigate a rather generic one-dimensional (1D) non-Hermitian Dirac model which contains a tachyon phase~\cite{Feinberg1967}. 
Such a non-Hermitian Hamiltonian has been recently realized experimentally on waveguide lattices on a photonic chip~\cite{Song2020, Xiao2021} as well as proposed as an effective Hamiltonian in an ion trap experiment~\cite{Lee2015}.
The system Hamiltonian in momentum space is 
\begin{equation}\label{eq:h_tachyon}
	H_0=\sum_{p} p\,c\,\sigma_x +\Delta\sigma_y-imc^2\sigma_z,
\end{equation}
which has two momentum-dependent energy bands $E_\pm(p)=\pm\e(p)$, with $\e(p)=\sqrt{p^2c^2+\Delta^2-m^2c^4}$ [see Fig.~\ref{fig:1}(a)], with the Fermi velocity as the effective light speed $c$.
The Hamiltonian is invariant with respect to a $\mathcal{PT}$-symmetry operator $\sigma_x\mathcal K$, with $\mathcal K$ performing complex conjugation, $[\sigma_x\mathcal K, H_0]=0$.
The gapped non-Hermitian Dirac Hamiltonian at $|\Delta|>mc^2$ is in a $\mathcal{PT}$-symmetric phase since the eigenvectors are invariant to the $\mathcal{PT}$-symmetry operator and the spectrum consists of two real energy bands~\cite{Bender1998,Bender2007,Mostafazadeh2002}.
In this regime, the density of states, $g(\omega)= \sum_p\delta(\omega-E_\pm(p))$ 
vanishes below the gap edge and features a sharp threshold 
singularity at $\sqrt{\Delta^2-m^2c^4}$.
Exactly at $|\Delta|=mc^2$, the spectrum becomes gapless with a linear dispersion $\e(p)=c|p|$.
The $p=0$ point marks an exceptional point (EP), where not only the eigenvalues, but also the eigenvectors coalesce~\cite{Berry2004,Heiss2012}.
Because of the linear dispersion, the density of states is constant for all energies.
For $|\Delta|<mc^2$, a tachyonlike spectrum with a hyperbolic gapless dispersion develops. 
Depending on $p$, the system is in a $\mathcal{PT}$-symmetric phase with real eigenvalues for $p^2c^2>m^2c^4-\Delta^2$ or in a $\mathcal{PT}$-symmetry broken phase, with conjugate pairs of imaginary eigenvalues, at $p^2c^2<m^2c^4-\Delta^2$.
These two momentum regions are separated by EPs at $\e(p_{\rm EP})=0$. 
Close to the EP, the spectrum behaves as $|\e(p)|\sim \sqrt{p-p_{\rm EP}}$, 
and consequently $g(\omega)\sim \omega$ for small energies. 
This is reminiscent of the case of 2D graphene~\cite{CastroNeto2009}, 
but in 1D such a behavior is only possible for non-Hermitian systems.  
Close to the EP, the group velocity diverges as $\pd_p \e(p)\sim 1/\sqrt{p-p_{\rm EP}}$.

As a perturbation, we consider a time-dependent  electric field $\mathcal E$ that couples to the current operator and 
induces optical transitions between the two bands.
The electric field effect is taken into account through a time-dependent vector potential $\mathcal A(t)$
as $p\to p+e\mathcal A(t)$, with $e$, the absolute value of electric charge, and $\mathcal E=-\pd \mathcal A/\pd t$. 
Then the perturbation couples to the vector potential through the current operator $j=-\delta H/\delta \mathcal A=\sum_p j_p$, with $j_p = -ec\sigma_x$.
In spite of being non-Hermitian, $H_0$ is still gauge invariant~\cite{Merzbacher1998,Benfatto2005}~\footnote{Gauge invariance in our case means that a time dependent electric field can equally be represented by a vector or a scalar potential}. 
This also implies that the local charge density $\avg{n(x)}$ is conserved and the continuity equation holds~\cite{SM}
\begin{equation}
	-e\pd_t \avg{n(x)} +\pd_x\avg{j(x)} = 0,\label{eq:continuity}
\end{equation}
which guarantees that the optical sum rule is satisfied, as we discuss later.

\paragraph{Optical conductivity.} 
Our main task is to determine the  current-current correlator $\chi(t,t')\equiv -\chi_{jj}(t,t')$ using Eq.~\eqref{eq:chi_general} and
the corresponding
optical conductivity, that describes the optical response to the external electric field.
As initial conditions we assume the half filling case.
In the gapped and $\mathcal{PT}$-symmetric tachyon phases, the system starts in the ground state with all the $-\e(p)$ levels initially occupied, while in the $\mathcal{PT}$-symmetry broken phase the levels with a larger weight, and corresponding to the slowest decay rate, associated with $+i|\e(p)|$ states dominate the dynamics at $t<0$~\cite{SM} are assumed to be filled. 

First of all, we find that while in general the perturbation causes transitions between the bands, 
the response is zero in the broken $\mathcal{PT}$-symmetry phase for tachyons, i.e.,~when the spectrum consists of pairs of imaginary eigenvalues.
This is due to exact cancellation between the generalized commutator contribution to the response, $[j_p(\tau), j_p(0)]$, and the norm corrections. 
We have also checked this numerically by solving Eq.~\eqref{eq:h_tachyon} in the presence of a weak electromagnetic field.
Therefore, only the $\mathcal{PT}$-symmetric part of the model with pairs of real eigenvalues contributes to linear response.
The initial conditions fix $\rho_0(t)=\rho(0)$, and Eq.~\eqref{chi_ps_h} becomes 
\begin{eqnarray}\label{eq:chi_r}
	\chi(\tau) &=&{\sum_p}'\frac{\chi(p,\tau)}{L}=\frac{i\theta(\tau)}{\hbar\, L}
	{\sum_{p}}'
	\avg{[j_p(\tau),j_p(0)]}_0\notag\\ &&-\avg{j_p(0)}_0\avg{[e^{iH_0^\dag\tau/\hbar}e^{-iH_0\tau/\hbar},j_p(0)]}_0,
\end{eqnarray}
where $L$ is the system size and $\sum_{p}'$ restricts the sum to momenta in the $\mathcal{PT}$-symmetric region $p^2c^2>m^2c^4-\Delta^2$.
For our particular model, at half filling, $\chi(\tau)$ is time-translation invariant, unlike the general formula in Eq.~\eqref{eq:chi_general}.

First we consider the case $\Delta=0$ with purely imaginary mass, where the salient features of non-Hermitian linear response are clearly manifested.
Noticing that the Hamiltonian $H_0(\Delta=0)$ is self-conjugate with respect to $\sigma_x$, it implies $[j_p, H_0(\Delta=0)]_\sim=0$, and therefore the current operator is time independent.
Hence the first contribution to $\chi(\tau)$ in Eq.~\eqref{eq:chi_r}, containing the commutator $[j_p(\tau),j_p(0)]= 0$, drops out. 
However, the norm correction term  remains finite and  contributes to the linear response.
Such a term is unique to non-Hermitian systems, without any analogue in the conventional Hermitian Kubo formula.

For generic real gap $\Delta$, both the generalized commutator and the norm correction terms contribute, and $\chi(p,\tau)$ reads
\begin{equation}\label{eq:corr_p}
	\chi(p,\tau) = \frac{2e^2c^2}{\hbar}\theta(\tau)\frac{(\e(p)\Delta-pmc^3)^2}{(p^2c^2+\Delta^2)^2}\sin\frac{2\e(p) \tau}{\hbar}.
\end{equation}
The result recovers correctly the Hermitian limit $m\to 0$~\cite{Dora2015,Lee1974}. 
The correlation function~\eqref{eq:corr_p} captures an interesting behavior for the Dirac-point dispersion $|\Delta|=mc^2$: only half of the momenta contribute to linear response. Indeed,
for $p>0$ and $\Delta=mc^2$,  $\chi(p>0,\tau)=0$ and $\avg{j_{p>0}(t)}=ec$, while for $p<0$ and $\Delta=-mc^2$, $\chi(p<0,\tau)=0$ and $\avg{j_{p<0}(t)}=-ec$.
This is a result of destructive quantum interference, due to the two masses of equal amplitude, that inhibits  half of the momentum states to transition from the ground to the excited states~\cite{SM}.
This is analogous to unidirectional invisibility in non-Hermitian systems~\cite{Lin2011,ElGanainy2018}.
Our construction is based on the assumption of a non-degenerate spectrum, so exactly at the EP the theory does not apply. Nevertheless, for any choice of parameters that brings the system close to the EP, the 
linear theory remains valid~\cite{SM}, and in particular, the correlation function vanishes close to the EP on both sides, which indicates that one would not expect unphysical anomalies due to the EP.

The time dependent $\chi(\tau)$ is obtained by summation over all momentum states. 
We find that in the gapped phase  $\chi(\tau)$ exhibits a damped oscillating behavior 
with a slow decay $\sim \tau^{-1/2}$ which transforms into an exponential decay for the 
linear dispersion and in the tachyon phase~\cite{SM}. 
The response in frequency space $\chi(\omega)$ follows by Fourier transforming Eq.~\eqref{eq:corr_p} and summing over momenta.
The real (absorptive) part of the optical conductivity is given by the imaginary part of the susceptibility, $\sigma'(\omega)=\chi''(\omega)/\omega$~\cite{SM}. 

Figure~\ref{fig:1}(b) displays the evolution of $\sigma'(\omega)$ in all three regimes.
For $|\Delta|=mc^2$, the dispersion is linear, similarly to the gapless 1D Dirac equation, but, in contrast to the expectations from a 1D Hamiltonian $\sim pc\sigma_x$, where $\sigma'(\omega)=0$~\cite{Lee1974}, now $\sigma'(\omega>0)>0$.
This is a manifestation of \textit{Zitterbewegung}, and follows from the noncommutativity of the current operator and Hamiltonian, due to the additional real and imaginary mass terms~\cite{Cserti2006, Thaller2011}.

The conductivity in the  gapped phase $|\Delta|>mc^2$, behaves qualitatively similar to 
the one in the gapped Hermitian model~\cite{Lee1974}.
There is a threshold for excitations given by the band gap $\hbar\omega\simeq 2\,\sqrt{\Delta^2-m^2c^4}$, above which
a threshold singularity  appears, similarly to the density of states.

Finally, in the tachyon phase $|\Delta|<mc^2$, $\sigma'(\omega)$ remains always finite,
with no anomalous behavior when the system approaches the EP, although the group velocity $\pd_p \e(p)$ diverges at the EPs. 
The low-frequency constant optical conductivity parallels closely to graphene, where the linear density of states produces similar behavior.
Furthering this analogy with graphene, we also identify a nonzero dc conductivity $\sigma_{\rm dc}=\sigma(\omega=0)$ for tachyons which depends on the ratio of the effective gap and the imaginary mass,
\begin{equation}
	\sigma_{\rm dc} = \frac{e^2}{4mc}\frac{|\sqrt{\Delta^2-m^2c^4}|}{mc^2},
\end{equation}
while $\sigma_{\rm dc } = 0$ for $|\Delta|\geq mc^2$. 
A finite dc conductivity without scatterers was only thought to be possible for graphene~\cite{Katsnelson2006,CastroNeto2009} in 2D,
and the present 1D non-Hermitian Dirac system represents another occurrence.
However, with increasing frequency, the analogy with graphene stops, as the EPs do not play a major role and the optical conductivity decays with frequency as $\sim \omega^{-3}$.
Finally, as a consequence of the  local charge conservation~\cite{SM, Benfatto2005}, the optical sum rule~\cite{Mahan2000} is satisfied for all phases of the non-Hermitian model,
\begin{equation}
	\int_0^\infty \sigma'(\omega)  d\omega = \frac{e^2c}{2\hbar}.
\end{equation}
It would be interesting to investigate this and other sum rules in non-Hermitian, higher dimensional systems as well.

\paragraph{Connection to experiments.} 
Our general theory is expected to find applications in a variety of non-Hermitian settings~\cite{Ashida2020}, where
additional external perturbations reveal various aspects of the underlying model. 
For example, the role of topology in non-Hermitian systems, combined with many-body physics, as well as various sources of dissipation can be probed using our theory.

In terms of the tachyon physics, imaginary mass particle dynamics has been recently demonstrated experimentally in waveguide lattices~\cite{Song2020}, and in single-photon 
interferometric devices~\cite{Xiao2021}, thus realizing the effective non-Hermitian Hamiltonian~\eqref{eq:h_tachyon}.
The waveguide lattices are composed by alternating waveguides where either loss or gain is dominant, and an alternating coupling coefficient between the propagating modes.
The mode propagation in the paraxial approximation is then mapped to a Dirac-like Hamiltonian where the imaginary mass results from the interplay 
between the ratio of coupling coefficients and gain/loss rates. 
Experiments on single-photon interferometry~\cite{Xiao2021} have already realized Eq.~\eqref{eq:h_tachyon}. Because of the extreme tunability of this setup, additional perturbations such as a
frequency dependent vector potential can be engineered, and the current can be measured separately for each $p$ mode. From this, $\chi(p,\tau)$ is determined, whose knowledge yields the optical
conductivity.

Another promising avenue involves ion trap physics, which allows us to simulate the 1D Dirac equation in a minimal setup consisting of two atomic levels and a motional degree of freedom~\cite{Lamata2007,Gerritsma2010,Lee2015}.
In such setups, the density matrix of the system evolves according to a Lindblad equation~\cite{Lindblad1976, Gorini1976}, but postselecting outcomes with no photon emission events singles out the dynamics driven by the non-Hermitian Hamiltonian in Eq.~\eqref{eq:h_tachyon}.
Finally, the optical conductivity can be measured by in situ fluorescence
spectroscopy~\cite{Anderson.2019}. 

\paragraph{Conclusions.}
To sum up, we developed a unified linear response theory for non-Hermitian systems and perturbations, and provided an expression for the Kubo formula suitable for non-Hermitian models. 
It contains (i) a generalized commutator, (ii) norm correction terms due to nonunitary dynamics, and (iii) is not generally time-translation invariant.
Through the generalized commutator, the non-Hermitian linear response
gives direct experimental access to unequal-time anticommutators of
observables instead of commutators, quantities hard to measure within the Hermitian realm.

We applied it to investigate the optical conductivity of a generic one-dimensional Dirac model, with both real and imaginary masses.
This model features a tachyon phase where excitations travel faster than an effective speed of light.
We find that the low-energy physics in this phase is graphenelike with constant optical conductivity and finite minimal conductivity.
For all masses, the optical sum rule is satisfied. 
We argue that these results can be tested experimentally.
Our Kubo formula represents an ideal starting point to gain information about non-Hermitian many-body systems as well, where the available methods are limited.

\paragraph{Acknowledgments.}
We would like to thank to  A. Mostafazadeh for clarifying discussions.
This research is supported by the National Research, Development and Innovation Office---NKFIH within the Quantum Technology National Excellence Program (Project No. 2017-1.2.1-NKP-2017-00001), K119442, K134437, and
by the Romanian National Authority for Scientific Research and Innovation, 
UEFISCDI, under Projects No.~PN-III-P4-ID-PCE-2020-0277 and No.~PN-III-P1-1.1-TE-2019-0423.

\bibliographystyle{apsrev4-2}
\bibliography{bibl}
\end{document}


\author{Doru Sticlet}
\email{doru.sticlet@itim-cj.ro}
\affiliation{National Institute for R\&D of Isotopic and Molecular Technologies, 67-103 Donat, 400293 Cluj-Napoca, Romania}
\author{Bal\'azs D\'ora}
\affiliation{MTA-BME Lend\"ulet Topology and Correlation Research Group, Budapest University of Technology and Economics, 1521 Budapest, Hungary}
\affiliation{Department of Theoretical Physics, Budapest University of Technology and Economics, 1521 Budapest, Hungary}
\author{C\u{a}t\u{a}lin Pa\c{s}cu Moca}
\affiliation{MTA-BME Quantum Dynamics and Correlations Research Group, Institute of Physics, Budapest University of Technology and Economics, 1521 Budapest, Hungary}
\affiliation{Department  of  Physics,  University  of  Oradea,  410087,  Oradea,  Romania}
\noaffiliation
	
\title{Supplementary Material for ``Kubo formula for non-Hermitian systems and tachyon optical conductivity''}

\maketitle

\section{Linear response theory}
For the sake of generality, we focus on a time-independent non-Hermitian Hamiltonian $H_0$ with nondegenerate energy eigenvalues perturbed by a time-dependent potential $V(t)$ which starts to act at $t=0$.
The time-dependent potential is $V(t) = Bf(t)$, where operator $B$ is a time-independent operator in the Schr\"odinger picture, not necessarily Hermitian. 
The external force $f(t)$ is considered without loss of generality as a real-valued quantity. 
The change in the expectation value $\avg{A(t)}$ of an operator $A$  due to the perturbation is encoded in the correlation function $\chi(t,t')$,
\begin{equation}\label{lin_resp}
\delta \avg{A(t)} = \int dt' \chi_{AB}(t,t')f(t').
\end{equation}

Expectation values are normalized by a time-dependent norm. 
Their dynamics is obtained using the equation of motion for the density matrix $i\hbar\pd_t\rho(t)=H\rho(t)-\rho(t)H^\dag$, where $H=H_0+V(t)$, which generates additional terms which occur only for non-Hermitian systems~\cite{Graefe2008},
\begin{equation}\label{expec_dyn}
i\hbar\pd_t\avg{A}=\avg{AH-H^\dag A}-\avg{A}\avg{H-H^\dag}.
\end{equation}
In order to treat the effect of $V(t)$, it is convenient to switch to the interaction picture, where operators evolve with $H_0$.
The conventional expression relating the interaction and the Schr\" odinger picture for the ket states remains the same, but since $H_0\ne H_0^{\dagger}$, the Hermitian conjugate expression for the bra changes, $\langle\psi_I(t)| = \langle\psi(t)| e^{-iH_0^\dag t/\hbar}$.

The evolution of an operator in the interaction representation is constructed in the conventional way, $ A_I(t) = e^{i H_0^\dag t/\hbar}A e^{-iH_0t/\hbar}$.
Since the evolution is not unitary, the density matrix in the interaction picture
\begin{equation}\label{interaction_picture}
\rho_I(t) = e^{iH_0t/\hbar}\rho(t)e^{-i H_0^\dag t/\hbar},
\end{equation}
is not properly normalized, and therefore $\avg{A(t)}$ also reads
\begin{equation}\label{avg_2}
\avg{A(t)}= \frac{\tr[\rho_I(t) A_I(t)]}{\tr[\rho(t)]},
\end{equation}
where the denominator is included for proper normalization.

Under the effect of the external perturbation, following the general strategy~\cite{Mahan2000}, we look at the changes in the density matrices in the 
Schr\"odinger and interaction pictures induced by the external perturbation. For that we write $\rho(t) = \rho_0(t)+\delta\rho(t)$ and $\rho_I(t) = \rho_{0,I}(t)+\delta\rho_I(t)$ where the density matrices $\rho_{0}$ evolve solely due to $H_0$, $\rho_0(t) = e^{-iH_0 t/\hbar}\rho(0) e^{iH_0^\dag t/\hbar}$.
Consequently, $\rho_{0,I}(t)$ is identified with the initial density matrix $\rho_{0,I}(t)=\rho(0)$.
The linear response requires estimating $\delta\rho$ only to first order in the perturbation. 
To this order, the operator expectation value~\eqref{avg_2} reads
\begin{equation}\label{del_A}
\avg{A}\simeq \avg{A}_0+\delta\avg{A},\quad 
\delta \avg{A(t)} = \frac{\tr[\delta\rho_I(t) A_I(t)]}{\tr[\rho_0(t)]}
- \avg{A(t)}_0\frac{\tr[\delta\rho(t)]}{\tr[\rho_0(t)]}.
\end{equation}
where  $\avg{A(t)}_0$ represents the expectation value in the absence of the perturbation
$\avg{A(t)}_0 = \tr[\rho_0(t)A]/\tr[\rho_0(t)]= \tr[\rho(0) A_I(t)]/\tr[\rho_0(t)]$.
Note that since $H_0$ is generally non-Hermitian, the norm of the density matrix under the evolution with $H_0$ may change in time, which requires an explicit time dependence in $\rho_0(t)$, even in the absence of $V(t)$.
To find an explicit expression for $\delta\rho_{I}$, we use the equation of motion for the density matrix in interaction picture
\begin{equation}\label{deriv_I}
i\hbar\frac{d\rho_I(t)}{dt} = e^{iH_0t/\hbar}e^{-iH_0^\dag t/\hbar} V_I(t)\rho_I(t) - \rho_I(t) V_I^\dag(t) e^{iH_0t/\hbar}e^{-iH_0^\dag t/\hbar},
\end{equation}
which gives the first-order correction to $\rho(0)$,
\begin{equation}\label{rho_I_lin}
\delta\rho_I(t)\simeq-\frac{i}{\hbar}\int_0^t dt'\big(e^{iH_0t'/\hbar}e^{-iH_0^\dag t'/\hbar}V_I(t')\rho(0) - \rho(0) V_I^\dag(t') e^{iH_0t'/\hbar}e^{-iH_0^\dag t'/\hbar}\big).
\end{equation}
The effect of the perturbation on the expectation $\avg{A(t)}$ is made explicit by using Eqs.~\eqref{interaction_picture} and~\eqref{rho_I_lin} into Eq.~\eqref{del_A}.
Then comparison with Eq.~\eqref{lin_resp} yields the generalized susceptibility
\begin{equation}\label{eq:chi_general}
\chi_{AB}(t,t')=-\frac{i}{\hbar}\theta(\tau)
\tr\bigg\{
\bigg[
[A(\tau), B]_\sim-\avg{A(t)}_0[e^{iH_0^\dag\tau/\hbar}e^{-iH_0\tau/\hbar},B]_\sim\bigg]
\cdot \frac{\rho_0(t')}{\tr[\rho_0(t)]}\bigg\},
\end{equation}
after the conventional identification, that
the operator time evolution in the interaction picture, $A_I(t) = e^{i H_0^\dag t/\hbar}A e^{-iH_0t/\hbar}$  is identical to the Heisenberg picture
time evolution in terms of $H_0$, i.e. $A(t)=A_I(t)$.

The rest of the section discusses some particular cases. 
For example, if the initial Hamiltonian $H_0$ is Hermitian, then the system density matrix remains time-independent, $\rho_0(t)=\rho(0)$ is usually just the equilibrium density matrix, and hence $\avg{A(t)}_0 = \avg{A(0)}_0$.
If the perturbation $B$ is Hermitian as well, the  classic Kubo formula follows immediately, 
\begin{equation}
\chi_{AB}(\tau)=-\frac{i}{\hbar}\theta(\tau)\avg{[A(\tau), B]}_0.
\end{equation}
Another case discussed in the literature corresponds to a Hermitian Hamiltonian $H_0$ perturbed with an anti-Hermitian potential $B^\dag=-B$.
Then general expression~\eqref{eq:chi_general} recovers the result derived in Ref.~\cite{Pan2020}, with additional norm corrections,
\begin{equation}
\chi_{AB}(\tau)=-\frac{i}{\hbar}\theta(\tau)\big(\avg{\{A(\tau), B\}}_0-2\avg{A}_0\avg{B}_0\big).
\end{equation}
If $H_0$ is some generic non-Hermitian Hamiltonian, but the perturbation $B$ is Hermitian, then one has to use Eq.~\eqref{eq:chi_general}, with the only simplification that the two modified commutators $[\dots]_\sim$ are replaced by regular commutators. 
Evidently, if the perturbation $B$ is anti-Hermitian, then $[\dots]_\sim$ are replaced by anti-commutators. Finally, note that the trace in Eq.~\eqref{eq:chi_general} must be taken over an orthonormal basis, and that eigenstates of $H_0$ are generally not orthogonal. This requires in general performing an additional orthogonalization step in order to take the trace.

Other simplifications occur for a typical $\mathcal{PT}$-symmetric model with real eigenspectrum, where the initial density matrix is constructed from the ground-state wavefunction $H_0|0\rangle =E_0|0\rangle $. 
In this case, $\rho(0)$ does not evolve in time, in the absence of $V(t)$,
\begin{equation}
\rho_0(t) = e^{-iH_0 t/\hbar}|0\rangle\langle 0|e^{iH_0^\dag t/\hbar} = \rho(0),
\end{equation}
since $E_0$ is real.
The correlation function becomes explicitly time-translation invariant and reduces to
\begin{equation}\label{chi_ps_h}
\chi_{AB}(\tau)= -\frac{i}{\hbar}\theta(\tau)
\avg{[A(\tau), B]_\sim
	- \avg{A(0)}_0 [e^{iH_0^\dag \tau/\hbar}
	e^{-iH_0 \tau/\hbar},B]_\sim}_0.
\end{equation}
The expression Eq.~\eqref{chi_ps_h} for the correlation functions for a $\mathcal{PT}$-symmetric model 
Hamiltonian remains valid and can be applied to any other pseudo-Hermitian Hamiltonian with a
real spectrum~\cite{Mostafazadeh2002,Zhang2019}.
The first term in the correlation function bears resemblance to the commutator in the Hermitian case. 
However, the second term has no obvious analogue, but its presence is crucial to obtain the correct linear response expression, as we discuss in the study of tachyons.

\section{The tachyon model and the optical conductivity}\label{sec:model}
The non-Hermitian tachyon model was introduced in Eq.~(5) in the main paper, but here we derive it as an effective Hamiltonian
for a Dirac particle with decay~\cite{Lee2015}. 
A relativistic particle is described by the Hamiltonian $H_p = pc\sigma_x+\Delta\sigma_y$ where $p$ is the momentum, and $\sigma_{x,y}$ are the regular Pauli matrices, and $\Delta $ is the energy gap. If the  excited levels of $H_p$ have a finite lifetime, modeled by a decay rate $\gamma$, the time evolution of the model is described by the  Lindblad equation for the density matrix, $\rho=\sum_{p}\rho_p$, with 
\begin{equation}
\frac{d\rho_p}{dt} = -\frac{i}{\hbar}[H_p,\rho_p] - \frac1 2\{L^\dag L, \rho_p\}+L\rho_p L^\dag,
\end{equation}
with Lindblad operators $L=\sqrt{\gamma}\sigma_-$, $\sigma_\pm = \sigma_x\pm i\sigma_y$. 
The non-Hermitian evolution is conditioned by the absence of photon emission events, i.e 
by neglecting the recycling (quantum jump) term $\sim L\rho L^\dag$.
Then the system is modeled by a non-Hermitian Hamiltonian 
$
H_0' = \sum_p H_p - \frac{i\hbar\gamma}{2}\sigma_+\sigma_- 
$
 which describes the dynamics. 
 Denoting $mc^2=\hbar\gamma/4$, apart from a term proportional to the identity matrix 
$\sigma_0$  we recover the Hamiltonian $H_0$ in the main text, 
\begin{equation}\label{eq:h_tachyon_SM}
H_0'= H_0+\delta H = \sum_{p} p\,c\,\sigma_x +\Delta\sigma_y-imc^2(\sigma_0+\sigma_z).
\end{equation}
The extra term, $\delta H = \sum_p-imc^2 \sigma_0$, shifts the eigenvalues in the complex plane by a constant $E_\pm =  -imc^2\pm \varepsilon$, with $\varepsilon = \sqrt{p^2c^2+\Delta^2-m^2c^4}$, but does not change the  eigenvectors. For a given momentum $p$ the eigenvectors are 
\begin{equation}\label{vecs}
	|p+\rangle = 
	\frac{1}{\sqrt{N}}
	\pmat{pc-i\Delta\\imc^2+\varepsilon
	},\quad 
	|p-\rangle 
	=\frac{1}{\sqrt{N}}
	\pmat{imc^2+\varepsilon\\-pc-i\Delta},
\end{equation}
where $|p+\rangle$ and $|p-\rangle$ correspond to $E_+$ and $E_-$ bands, respectively.
The normalization factor depends on the phase in which the system is investigated: $N=2(\e^2+m^2c^4)$ if $\e$ is real, and $N=2mc^2(|\e|+mc^2)$ if $\e$ is imaginary.

The constant shift term in the effective Hamiltonian $\delta H$ is required in the Lindbladian evolution as it 
ensures that all eigenvalues of the Lindbladian have negative imaginary part, and the dynamics evolves corectly 
towards the steady state. 
It implies as well that all the eigenstates of $H_0'$ display the same exponential decay. 
Still, when computing any expectation value $\avg{A}_0=\tr[\rho_0(t)A]/\tr[\rho_0(t)]$, the result is unaffected by the constant decay in the wave function due to the explicit normalization term in the denominator, that cancels exactly the decay in the numerator.
This also implies that the correlation 
functions that we evaluate are invariant with respect to the presence or absence of the constant term $\delta H$ in the Hamiltonian. 
Given this, the general Kubo formula [Eq. (4) in the main manuscript], yields the same answer with or without the constant imaginary term included. 
In the main manuscript, for sake of clarity, we focused on the simplified Hamiltonian $H_0$, and we have not included the constant imaginary term that breaks the $\mathcal{PT}$-symmetry.

By construction, the Hamiltonian $H_0$ is invariant under a $\mathcal{PT}$-type symmetry represented by the antiunitary operator $\sigma_x\mathcal K$, where $\mathcal K$ performs complex conjugation, $[\sigma_x\mathcal K,H_0]=0$. 
The model is in a $\mathcal{PT}$-symmetric phase when also the eigenvectors are invariant to the symmetry $\sigma_x\mathcal K|p\pm\rangle =|p\pm\rangle$ and consequently the eigenvalues are real. In the $\mathcal{PT}$-symmetry broken phase $\sigma_x\mathcal K|p\pm\rangle =|p\mp\rangle$ and the eigenvalues are imaginary.

The effect of an electric field perturbation is incorporated using the minimal coupling scheme $p\to p+e\mathcal A(t)$, with  $\mathcal A(t)$, 
the associated vector potential and $e$, the absolute value of the electric charge.
Consequently the full Hamiltonian reads $H = H_0+V(t)$, with $V(t)=\sum_pec\mathcal A(t) \sigma_x$.  Correspondingly, the current operator is 
\begin{equation}
j = -\delta H/\delta\mathcal{A}=-\sum_p ec\sigma_x.\label{eq:current}
\end{equation}

In spite of being non-Hermitian, the Hamiltonian $H_0$ remains gauge invariant in the sense that a time-dependent electric field can equally be represented by a vector or scalar potential. 
This is equivalent to local charge conservation, and the conventional continuity equation holds.
This would also imply that the $f$-sum rule for the optical conductivity is expected to be satisfied~\cite{Benfatto2005}.
It is convenient to prove the continuity equation $-e\avg{n(x)}+\pd_x\avg{j(x)}=0$ using second quantization, with expectation values taken in the ground state, and properly normalized by the respective wave function.
The local charge density and its Fourier transform read as
\begin{equation}
-en(x)=-\frac{e}{L}\sum_q n_q e^{iqx},\quad
n_q=\sum_{k\alpha} c^\dag_{k\alpha}c_{k+q\alpha},
\end{equation}
with $c_{k\alpha}$, a fermion annihilation operator indexed by momentum $k$ and spin $\alpha$.
The local charge density evolution is determined using Eq.~\eqref{expec_dyn} from the behavior of the $n_q$ component,
\begin{eqnarray}\label{cont0}
-e\pd_t \avg{n_q} &=& \frac{ie}{\hbar}\big(\avg{n_q\mathcal H_0-\mathcal H_0^\dag n_q}-\avg{n_q}\avg{\mathcal H_0-\mathcal H^\dag_0}),\notag\\
&=&\frac{ie}{\hbar}(\avg{[n_q,\mathcal{H_+}]}+\avg{\{n_q,\mathcal{H_-}\}})-2\avg{n_q}\avg{\mathcal H_-},
\end{eqnarray}
with $\mathcal H_0=\sum_{k\alpha\beta}c^\dag_{k\alpha}H_0^{\alpha\beta}(k)c_{k\beta}$, the translation invariant Hamiltonian of the tachyon model.
We have decomposed for convenience the Hamiltonian into its Hermitian $(+)$ and anti-Hermitian $(-)$ parts, 
\begin{equation}
\mathcal H_{\pm} = (\mathcal H_0\pm \mathcal H_0^\dag)/2.
\end{equation}
The commutator expectation value is identified with the current density expectation value 
\begin{eqnarray}\label{simp}
\frac{ie}{\hbar}\avg{[n_q,\mathcal{H_+}]} &=& \frac{ie}{\hbar}
\sum_{k\alpha\beta} [H_+^{\alpha\beta}(k+q)-H_+^{\alpha\beta}(k)]\avg{c^\dag_{k\alpha}c_{k+q\beta}}
= ie\sum_{k\alpha\beta} q \avg{c^\dag_{k\alpha}\sigma_{x}^{\alpha\beta}c_{k+q\beta}}
\notag\\
&=&-iq\avg{j_q},
\end{eqnarray}
with $H_+(k)=\hbar k c\sigma_x+\Delta\sigma_y$. Eq.~\eqref{simp} holds in general for arbitrary quadratic and time-translation invariant Hamiltonians in the long wavelength limit (small $q$), and it is exact in the present case, for Hamiltonians linear in momentum. 
Note that $j_q$ thus obtained is exactly the second quantized version of the one in the minimal coupling scheme Eq.~\eqref{eq:current}.
Then the local current density is reconstructed as $j(x)=\sum_q j_q\exp(iqx)/L$.
Back in Eq.~\eqref{cont0}, we find that the continuity equation reduces to
\begin{equation}\label{cont1}
-e\pd_t \avg{n_q} + iq\avg{j_q} = \frac{ie}{\hbar}(\avg{\{n_q,\mathcal H_-\}} - 2\avg{n_q}\avg{\mathcal H_-}),
\end{equation}
supplemented on the right hand side with the source terms, which arise from the interaction with the environment.
We demonstrate below that these also vanish for the current model.

The first source term reads
\begin{equation}\label{rhs1}
\avg{\{n_q,\mathcal H_-\}} =
\sum_{kk'\alpha\beta\alpha'}
\left\langle c^\dag_{k\alpha}H_-^{\alpha\beta}(k)c_{k\beta}c^\dag_{k'\alpha'}c_{k'+q\alpha'}
\right\rangle+
\left\langle
c^\dag_{k'\alpha'}c_{k'+q\alpha'}c^\dag_{k\alpha}H_-^{\alpha\beta}(k)c_{k\beta}\right\rangle,
\end{equation}
and here $H_-(k) = -imc^2\sigma_z$. This is only non-vanishing for $q=0$ due to the model being both non-interacting, with the ground-state wave function, a Slater determinant, and translation invariant. 
In this case, the operator $\sum_{\alpha'}c^\dag_{k'\alpha'}c_{k'\alpha'}$ is the total number of particles
in momentum state $k'$ and commutes with both $H_0$ and $H_0^\dag$. 
Consequently, the wave function is an eigenstate of this operator and Eq.~\eqref{rhs1} is equal to $2\avg{\mathcal H_-}N$, with $N=\avg{n_{q=0}}$, the total number of fermions in the system.
This is compensated exactly by the second source term of Eq.~\eqref{cont1} after noting again that $\avg{n_q}=\avg{n_{q=0}}=N$. 
Therefore, the source terms vanish and, after Fourier transforming back to real space, one obtains the conventional continuity equation,
\begin{gather}\label{cont2}
-e\pd_t \avg{n(x)} +\pd_x\avg{j(x)} = 0.
\end{gather}

\begin{figure}[t]
\includegraphics[width=\textwidth]{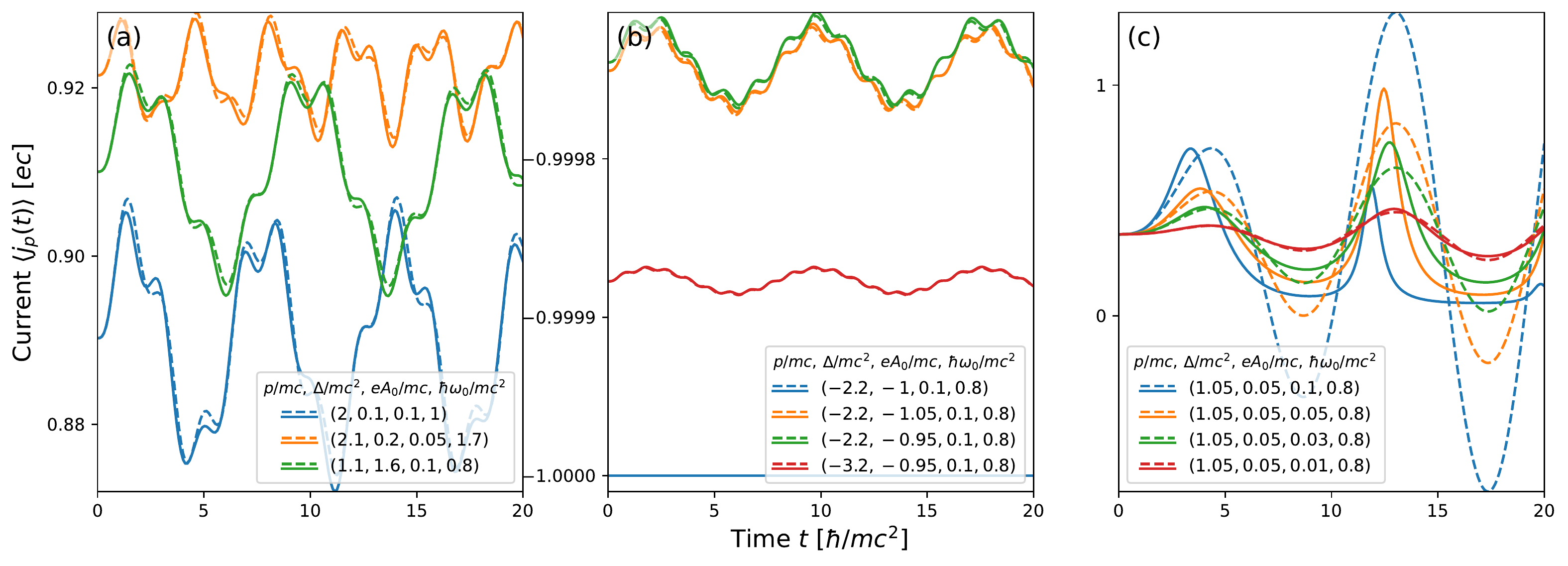}
\caption{Comparison between the current operator $\avg{j_p(t)}$ from the numerical solution to the  time-dependent Schr\"odinger equation (dashed lines) and the linear response approach (solid lines).
The legend shows the parameters sets at different momenta $p$, masses $\Delta$, vector potential amplitudes $A_0$, and frequencies $\omega_0$. 
Expectation values at (a) some generic points in parameter space, (b) near the critical point $\Delta=-mc^2$ (similar to $\Delta=mc^2$), and (c) near the exceptional point $\e(p)=0$.}
\label{fig:comp}
\end{figure}

The optical conductivity gives the current response to the external perturbation $V(t)$ and requires the calculation of the  current-current correlation function. In order to compare the analytical results with the numerical data we consider 
the case of a monochromatic perturbation, with an oscillating behavior for the vector potential of the form
$\mathcal A(t) = A_0\sin(\omega_0 t)$.
Fig.~\ref{fig:comp} shows a comparison between the linear response results and the numerical ones. The latter are obtained by integrating  the time-dependent 
Schr\" odinger equation and computing the averaged current $\avg{j_p(t)}$ for  different sets of parameters, as indicated in the legend.

The  linear response theory assumes that the perturbation is much smaller compared to the energy separation between the eigenvalues, and  
within this approximation it reproduces almost perfectly the numerical results in Fig.~\ref{fig:comp}(a) and (b).
The approximation only starts to fail very close to the exceptional point, but is improved by decreasing the perturbation amplitude [see Fig.~\ref{fig:comp}(c)].

For the Dirac-point dispersion, when $\Delta=mc^2$, the correlator $\chi(p, \tau)$ vanishes for positive momenta, while at $\Delta=-mc^2$, for negative momenta. 
In these cases the current expectations value is determined entirely by $\avg{j_p(0)}_0=ec$, for $\Delta=mc^2$, and $\avg{j_p(0)}_0=-ec$, for $\Delta=-mc^2$.
The latter case is illustrated both in the numerical and analytical calculation in Fig.~\ref{fig:comp}(b) (blue line). 
To understand this effect it is relevant to study the transition amplitudes between the current eigenstates for the Hamiltonian at $\Delta=mc^2$, $H_0(p)=pc\sigma_x+\Delta\sigma_y-i\Delta\sigma_z$.
Under a unitary transformation $U=\exp(i\pi(\sigma_x+\sigma_y+\sigma_z)/3\sqrt3)$ the Hamiltonian reads 
\begin{equation}
UH_0(p)U^\dag = \pmat{pc & 0 \\ 2\Delta & -pc}.
\end{equation}
For $p>0$,  $|p-\rangle=(0,1)^T$ belongs to the ground state and $|p+\rangle$ to the excited spectrum. 
The system cannot be excited, but can only decay
with a decay rate $\sim 2\Delta$. 
Then in the $T=0$ limit, when all $|p-\rangle$ states are filled, the system does not respond to the electric field at $p>0$ and the current remains independent in time, i.e.  $\avg{j_{p>0}(t)}=-ec\langle p-|\sigma_z |p-\rangle=ec$. 
At $p<0$, the lowest energy state is $|p-\rangle=(1,0)^T$, and now the system can be excited with a 
 $2\Delta$ rate. In conclusion, only a half of all momentum states in the system respond to the electric field.
Similar considerations work for $\Delta=-mc^2$, where $H_0=pc\sigma_x+\Delta\sigma_y+i\Delta\sigma_z$, and now $\avg{j_{p<0}(t)}=-ec$, with transitions at $p<0$ inhibited.  

\begin{figure}[t]
\includegraphics[width=0.4\textwidth]{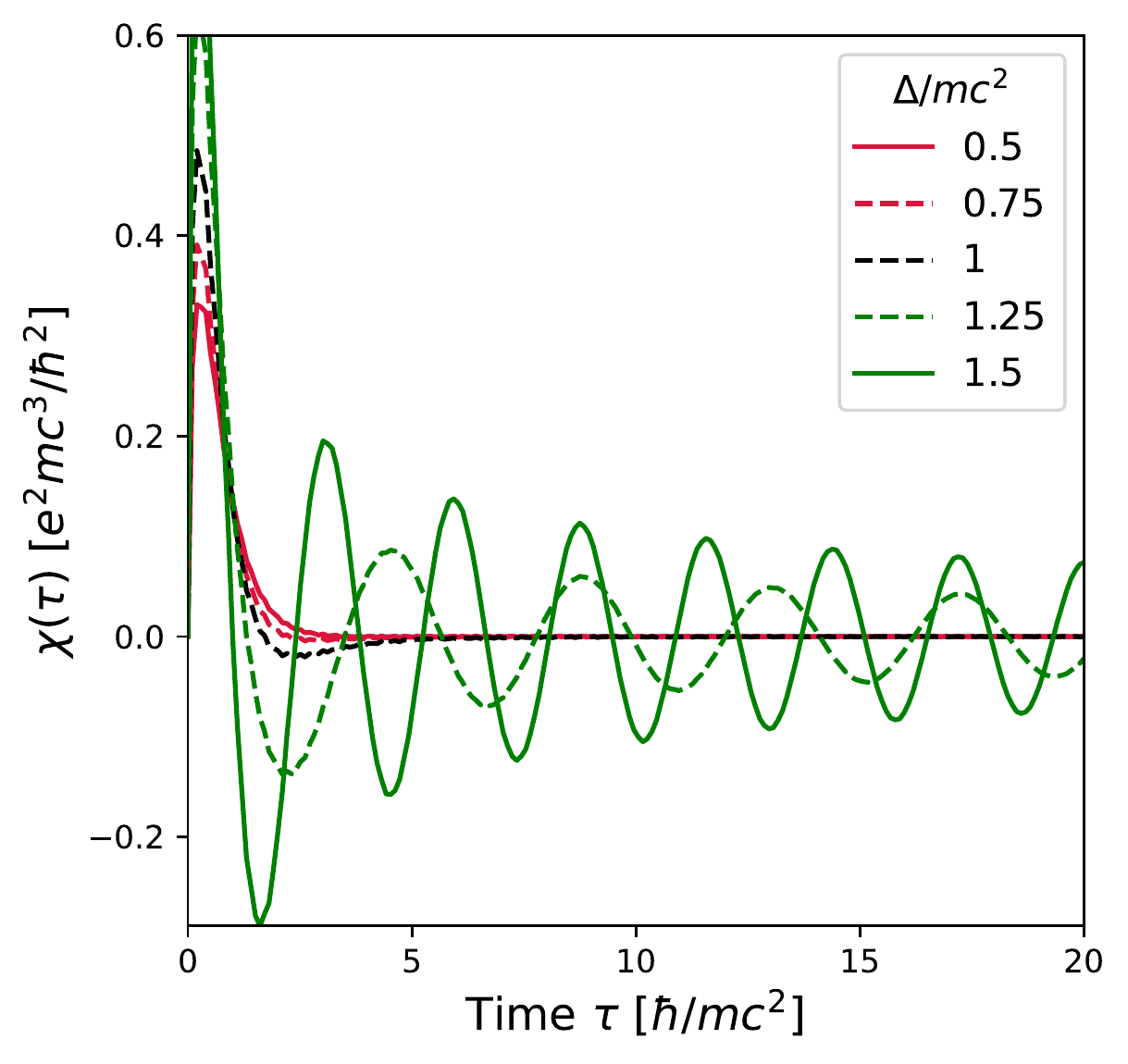}
\caption{Numerical results for current-current correlations as a function of time match the asymptotic behavior from Eq.~\eqref{chit_asymp}. There is an exponential decay for the gapless spectrum $|\Delta|\le mc^2$ and a damped oscillatory behavior for
$|\Delta|>mc^2$.}
\label{fig:chi_t}
\end{figure}

The correlation function integrated over momenta gives the time-dependent susceptibility, $\chi(\tau) = \int dp\chi(p,\tau)/2\pi\hbar.$
An asymptotic analysis, using the saddle-point approximation for the integrals, determines the time-dependence of correlations described in the main text,
\begin{equation}\label{chit_asymp}
\chi(\tau)\sim
\begin{cases}
-\frac{8e^2 c}{3\pi\hbar^3} m^2c^4\tau e^{-2mc^2\tau/\hbar},& |\Delta|=mc^2\\
-\frac{e^2c}{\sqrt{\pi\hbar^3\tau}}\frac{|\Delta_{\rm eff}|^{5/2}}{\Delta^2}
e^{-2|\Delta_{\rm eff}|\tau/\hbar},& |\Delta|<mc^2\\
\frac{e^2c}{\sqrt{\pi \hbar^3\tau}}
\frac{\Delta_{\rm eff}^{5/2}}{\Delta^2}
\sin(2\Delta_{\rm eff}\tau/\hbar + \pi/4),& |\Delta|>mc^2,
\end{cases}
\end{equation}
with $\Delta_{\rm eff}=\sqrt{\Delta^2-m^2c^4}$, the effective band gap.
Fig.~\ref{fig:chi_t} displays the numerical results for $\chi(\tau)$ in all these phases, and confirms the above asymptotic results.
A Fourier transform of $\chi(p,\tau)$ readily gives $\chi(\omega,p)$ and integration over momentum yields the total correlation function $\chi(\omega)=\int dp \chi(\omega,p)/2\pi\hbar$.
The absorptive part of the spectrum  is related to the imaginary part of the current-current correlation function, $\sigma'(\omega)=\chi''(\omega)/\omega$.
The dc conductivity is $\sigma_{\rm dc}\equiv\sigma(0)=\sigma'(0)$, since $\sigma''(-\omega)=-\sigma''(\omega)$.

In the gapped phase, $|\Delta|>mc^2$, the dc conductivity is trivially zero and $\chi''(\omega)$ reads
\begin{equation}\label{chi_gap}
\chi''(\omega) = \frac{e^2c}{2\hbar}\frac{\bar\omega\Delta^2_{\rm eff}
	}{\sqrt{\bar\omega^2-1}}
\frac{\bar\omega^2(\Delta^2+m^2c^4)-m^2c^4}{
(\bar \omega^2\Delta_{\rm eff}^2+m^2c^4)^2}
\theta(|\bar\omega|-1),
\quad
\bar\omega = \frac{\hbar\omega}{2\Delta_{\rm eff}}.
\end{equation}
For the linear dispersion $|\Delta|=mc^2$, 
\begin{equation}\label{chi_lin}
\chi''(\omega) = \frac{e^2c}{\hbar}\frac{\bar\omega^2}{(1+\bar\omega^2)^2}\text{sgn}(\bar\omega),
\quad
\bar\omega=\frac{\hbar\omega}{2mc^2},
\quad 
\lim_{\omega\to 0}\sigma'(\omega)=\lim_{\omega\to 0}\frac{e^2\hbar|\omega|}{4m^2c^3},
\end{equation}
 and the  dc conductivity vanishes. In the tachyon phase $|\Delta|<mc^2$, $\chi''(\omega)$ and the finite dc conductivity are
\begin{equation}\label{chi_tach}
\chi''(\omega) = \frac{e^2c}{2\hbar} \frac{\bar\omega|\Delta_{\rm eff}|^2}{\sqrt{\bar\omega^2+1}}
\frac{\bar\omega^2(\Delta^2+m^2c^4)+m^2c^4}{(\bar\omega^2|\Delta_{\rm eff}|^2+m^2c^4)^2},
\quad
\bar\omega=\frac{\hbar\omega}{2|\Delta_{\rm eff}|},\quad
\sigma_{\rm dc}
= \frac{e^2}{4mc}\frac{|\Delta_{\rm eff}|}{mc^2}.
\end{equation}
The following optical sum rule indicates the conservation of the total spectral weight for conductivity $\int_0^\infty d\omega \sigma'(\omega)  = e^2c/2\hbar$
in all the phases of the model. 
In general, the optical sum-rule is a consequence of the charge conservation, so the conservation of the spectral weight comes natural following Eq.~\eqref{cont2}.

\bibliographystyle{apsrev4-2}
\bibliography{bibl}